# Quantum Cryptography Using Various Reversible Quantum Logic Gates in WSNs

S. Ahmed[1], N. Javaid[2], S. H. Bouk[2], A. Javaid[3], M. A. Khan[2], Z. A. Khan[4]

[1]*Abasyn University, Peshawar, Pakistan*

*COMSATS Institute of Information Technology,* [2]*Islamabad,* [3]*Wah Cantt, Pakistan*

[4]*Faculty of Engineering, Dalhousie University, Halifax, Canada*

*Abstract*

**As sensor nodes are deployed anywhere in a wireless sensor network, hence their communication can be easily monitored. In these networks, message protection and node identification are very issues. Hence, security of large scale such networks requires efficient key distribution and management mechanisms. Quantum cryptography and particularly quantum key distribution is such a technique that allocates secure keys only for short distances. While not completely secure, it offers huge advantages over traditional methods by the use of entanglement swapping and quantum teleportation. Reversible logic gates like CNOT, Toffoli, Fredkin etc. are of basic importance in Quantum Computing. In our research, we adopted a EPR-pair allocation scheme in terms of these quantum gates to overcome the susceptibility caused by malicious nodes. As the qubits stored in a sensor node can be used only once and cannot be duplicated, hence risk of information leakage reduced even if the node are compromised.**

## I. INTRODUCTION

Wireless sensor networks (WSNs) has seen the expansion of certain opportunities and services in recent years with the growth of emerging wireless communication technologies. These networks are comprised of vastly scattered lightweight sensor nodes, and can be deployed in large quantity without any specific topology for the observation of some environment or system. As these nodes are normally operated by limited power and other constraints, WSNs come across strict resource limitations.

Sensor nodes perform three major functions of sensing, local data processing on sensed data, and communication of the data to neighboring nodes. Networks of such nodes give rise to a more trustful and accurate network, although each sensor is limited in its sensing ability, energy level,and processing power. Sensors collaborate and cooperate with each other, elect their cluster heads, aggregate their data and then transmit more refined results to a central location called the Base Station (BS). Major functions performed by the network require energy-efficient design and processed and concise information be delivered to BS.

As it is widely accepted that quantum cryptography offers quite improved security for communication over classical cryptography, hence, we adopted the quantum characteristics in the wireless sensor network to enhance the security. The qubits stored in the sensor node cannot be duplicated and can be used only once, the risk of information leakage is decreased even if the node is compromised [3]. Since nodes are deployed mostly in malicious environment, the security demand is critically important for WSNs. Assumption of the identity of some other node or modification of critical information by capturing sensors are damaging attacks and need to be avoided. Key assignment to sensors is regarded as the fundamental and critical issue in the security of wireless sensor networks. In the previous studied schemes, the authors did not apply any scheme for distributed quantum wireless network, and in paper [3] only CNOT gate was used and the security application becomes more strong when such quantum gates are applied which have multiple inputs and multiple outputs. Certain key management techniques in WSNs have been proposed for sensor networks [1].

Security in WSN has six major challenges[2]:

i. Wireless nature of communication,
ii. Limitation of resources on sensor nodes,
iii. Large and dense network,
iv. Lackage of fixed infrastructure,
v. No identified network topology before deployment,
vi. Risk of physical attacks.

The scientific contributions of this research paper are:

i. Security of large scale wireless networks requiring efficient key distribution and management mechanisms
ii. Quantum key distribution offers huge advantages over traditional methods by the use of entanglement swapping and quantum teleportation
iii. Adoption of a EPR-pair allocation scheme in terms of quantum gates like Toffoli and Feynman to overcome the susceptibility caused by malicious nodes.

As classical computation has the ability to store and operate on information on combination of bits, quantum computation works on quantum systems called qubits. In contrast to classical bit, which has either of the two mutually exclusive states, a qubit exists in a superposition of two states. The most general model of a qubit is:

$$|\psi\rangle = \alpha|0\rangle + \beta|1\rangle$$

where $\alpha$ and $\beta$ are complex numbers such that:

$$|\alpha|^2 + |\beta|^2 = 1 \quad and$$

$$|0\rangle = [1, 0]^T, |1\rangle = [1, 0]^T$$

are orthonormal basis vectors in 2-dimensional state space of a qubit. $\alpha$ and $\beta$ have the physical meanings that any given measurement of qubit shows the system to be in state $|0\rangle$ with probability $|\alpha|^2$ and in state $|1\rangle$ with a probability of $|\beta|^2$.

Qubits can be entangled with other qubits which is the real reason for the surprising computational power of a quantum computer. Entanglement is a uniquely quantum phenomenon, defined as a property of a multi-qubit state space to be a resource that the measurement on one qubit directly affects the other [4]. Process of extracting information from a set of qubits is called measurement. Entanglement takes place when particles such as electrons, photons, molecules interact physically with each other and then become separated; and the interaction is such that each resulting member of a pair is properly described by the same quantum mechanical state, which is indefinite in terms of position, momentum, spin, polarization, etc.

Quantum entanglement is a form of quantum superposition. After a measurement is made causing one member of such a pair to take a definite value (like clockwise spin), then the other member of entangled pair is found to have correctly taken the correlated value (that is the counterclockwise spin). Hence, a correlation exists between the measurements performed on entangled pairs, and this correlation is still observed even the pair is departed by large distances.

Quantum cryptography offers enhanced security for communication in comparison to classical one. In this research the quantum characteristics in WSN are considered to enhance the security. The research focuses on the allocation of EPR-Einstein Podolsky-Rosen pairs to each node which can transmit quantum information to another node through the mechanism of quantum teleportation, entrusting that the information transmitted will not be not eavesdropped. As the qubits stored in the sensors cannot be duplicated and can be used only once, hence risk of information leakage reduces to a larger extent even if the node is compromised [3].

The remainder of the paper is organized as follows: In section II, we have highlighted the related work of quantum cryptography in the field of wireless networks and the motivation behind that. Section III presents the proposed seurity model in terms of Toffoli gate and Feynman gate and their mathematical implementations, and finally section IV identifies the conclusion and the future work in the direction of the research.

## II. RELATED WORK AND MOTIVATION

In sensor networks, use of a single shared key is not a promising approach because a rival can easily obtain and break the key. Sensor nodes have to therefore, acclimatize their environments to develop a secure network by:

i. The use of pre-distributed keys
ii. Exchange of information only with the immediate neighbors, or
iii. Exchange of information with computationally robust nodes [1].

The field of quantum cryptography has the advantage of unique and unusual behavior of very small particles to enable users to develop their own secret keys alongwith detecting eavesdropping. Regular work on quantum cryptography began in late 1960's by Stephen J. Wiesner, but the first protocol for transmission of a private key using quantum techniques came on screen by 1984 by Bennett and Brassard. The development in the arena of quantum cryptography was motivated by the short-comings of classical cryptographic methods. Although, the encrypting and decrypting keys are different, it is not essential to distribute a key securely [3]. Further developments were made by [Malan et al. 2004; Gaubatz et al. 2004; Huang et al. 2003] to customize public key cryptography and elliptic key cryptography for low-power devices, but such approaches are still considered ineffecient due to their high processing requirements [1].

Quantum cryptography solves the problems of secret-key sharing by providing a mechanism for two users who are in different locations for the establishment of a secure secret key and detection of eavesdropping if occurred[3].

A researcher Cheng et al. [4] introduced the mechanism of quantum routing in sensor networks by the use of quantum teleportation and quantum logic gates. Nguyen et al. [5] discussed the integration of quantum cryptography with 802.11i security mechanisms and Lin et al. [6] proposed a scheme for the prevention of channel attacks and detection of wicked sensors by sharing a quantum table between sender and receiver before transmission. Huang et al. [7] presented a framework combining QKD and the wireless security standard. These proposals opened the doors to advancement in quantum networks.

These studies give an inspiration for the design of a framework of quantum WSNs with more vigorous security than conventional schemes with the help of quantum teleportation. This research presents an EPR pair management and allocation scheme in a quantum WSN to sensor nodes before they are randomly deployed. There is no need for central management and global information sharing and each node communicates with others through quantum teleportation. A sample scheme for the implementation of security for wireless sensor networks using quantum logic gates can be represented by Fig. 1.

Quantum reversible logic design attracted much attention in low-power design, optical computing, quantum cryptography and quantum computing. A network for reversible logic consists of a cascade of reversible gates [8].

A quantum reversible logic circuit must satisfy the following features [9]:

- Minimum number of reversible gates utility
- Minimum number of garbage outputs desired
- Minimum constant inputs be used

Such type of gates have been designed which maintain all information passed to them, so that the computation can be run in both forward and backward directions. The computation results in a large amount of junk, as each intermediate step is to be remembered, but heat generation is eliminated during the course of computation. A quantum logic gate is a basic circuit operating on a small number of qubits. Quantum logic gates are reversible, unlike many classical logic gates and are represented by unitary matrices. Some universal classical logic gates, like Toffoli gate, provide reversibility and can be directly mapped onto quantum logic gates.

One such gate, designed by Toffoli and Fredkin, known as Toffoli gate, is represented diagrammatically in Fig.1 below.

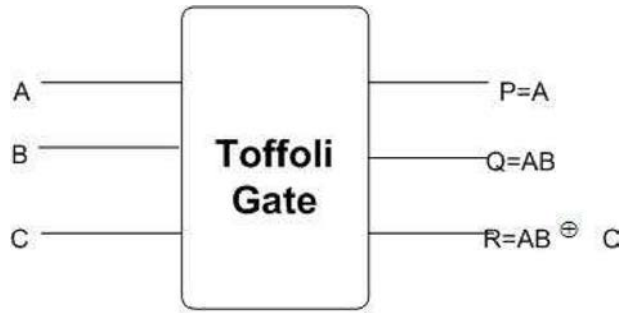

Fig. 1. 3x3 Toffoli gate

The output of this gate may be decomposed as [10]:

$$B \oplus (A,C) = \begin{cases} A.C & for\ B=0 & (AND) \\ A \oplus B & for\ C=1 & (XOR) \\ \overline{A} & for\ B=C=1 & (NOT) \end{cases}$$

where A.B represents an AND gate, $A \oplus B$ represents an XOR gate and $\overline{A}$ represents a NOT gate. This gate is regarded as universal, because it performs the functions of AND, XOR, or NOT depending on its inputs. Combination of many such gates can then be used for any computation and would still be reversible [10].

Another logic gate proposed by Feynman is a 2x2 reversible quantum logic gate which is represented as in the Fig.2 below which produces the following outputs if it has the inputs of A and B.

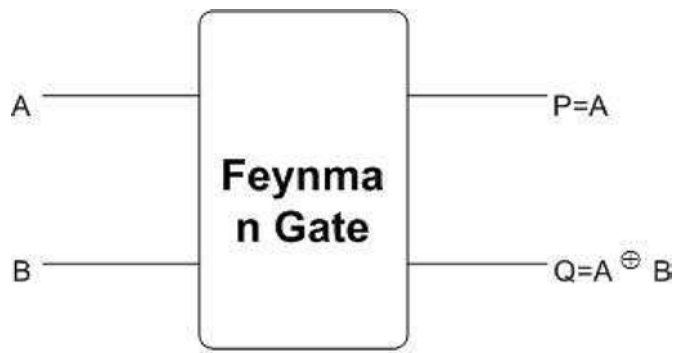

Fig. 2. 2x2 Feynman gate

### III. PROPOSED SECURITY MODEL

Most considerable usage of quantum entanglement is in the area of quantum teleportation. Bennett et al. [11] in 1993, first proposed a quantum technique for teleportation. His protocol implements intangible transfer of states of a system to a remote system. Unidentified quantum state of one qubit can be securely transported through the use of EPR pairs from sender to receiver by performing local operations and normal transmission methods. The arena of quantum teleportation has many applications, like transmission of quantum states in noisy environments and sharing states in distributed networks. Previously this work was done by [12] in which quantum teleportation was implemented by CNOT gate but in this research we have implemented quantum teleportation on a 2x2 Feynman gate in which the mathematical steps involved are more complex and hence more secure than CNOT gate. Again the same procedure is adopted on a 3x3 Toffoli gate to make it more secure and the final analytical results received by both Feynman and Toffoli gates are computationally identical to those received by CNOT in [28] but comparatively more secure.

First applying the procedure using 2x2 Feynman gate.

As in (1), a quantum state $|\phi_0\rangle$ is initially formed by the tensor product of three separate states that are kept by sender and receiver respectively.

$$|\phi_0\rangle = |\phi\rangle \otimes |0\rangle \otimes |0\rangle$$

$$|\phi_0\rangle = (\alpha|0\rangle + \beta|1\rangle) \otimes |00\rangle$$

$$|\phi_0\rangle = \alpha|000\rangle + \beta|100\rangle \tag{1}$$

Then a Hadamard gate is applied to the second qubit and quantum state $|\phi_1\rangle$ is achieved

$$|\phi_1\rangle = \alpha|0\rangle \otimes \frac{1}{\sqrt{2}}(|0\rangle + |1\rangle) \otimes |0\rangle + \beta|1\rangle \otimes \frac{1}{\sqrt{2}}(|0\rangle + |1\rangle) \otimes |0\rangle$$

$$|\phi_1\rangle = \frac{\alpha}{\sqrt{2}}(|000\rangle + |010\rangle) + \frac{\beta}{\sqrt{2}}(|101\rangle + |110\rangle) \tag{2}$$

Next, the 2x2 Feynman gate is applied from the second to the third qubit as equation (3) presents

$$|\phi_2\rangle = \frac{\alpha}{\sqrt{2}}(|0\rangle + FG(|00\rangle) + |0\rangle + FG(|10\rangle)) + \frac{\beta}{\sqrt{2}}(|1\rangle + FG(|01\rangle) + |1\rangle + FG(|10\rangle))$$

$$|\phi_2\rangle = \frac{\alpha}{\sqrt{2}}(|000\rangle + |011\rangle) + \frac{\beta}{\sqrt{2}}(|101\rangle + |111\rangle) \tag{3}$$

Again, the Feynman gate is applied from the first to the second qubit, which transforms $|\phi_2\rangle$ into $|\phi_3\rangle$

$$|\phi_3\rangle = \frac{\alpha}{\sqrt{2}}(FG|00\rangle \otimes |0\rangle + FG|01\rangle \otimes |1\rangle) + \frac{\beta}{\sqrt{2}}(|FG|10\rangle \otimes |1\rangle + FG(|11\rangle \otimes |1\rangle)$$

$$|\phi_3\rangle = \frac{\alpha}{\sqrt{2}}(|000\rangle + |011\rangle) + \frac{\beta}{\sqrt{2}}(|111\rangle + |101\rangle) \tag{4}$$

Now if the Hadamard gate is applied to the first qubit it produces $|\phi_4\rangle$

$$|\phi_4\rangle = \frac{\alpha}{\sqrt{2}} \times \frac{1}{\sqrt{2}}(|0\rangle + |1\rangle) \otimes (|00\rangle + |11\rangle) + \frac{\beta}{\sqrt{2}} \times \frac{1}{\sqrt{2}}(|0\rangle - |1\rangle) \otimes (|11\rangle + |01\rangle)$$

$$|\phi_4\rangle = \frac{1}{2}|00\rangle \otimes (\alpha|0\rangle + \beta|1\rangle) + \frac{1}{2}|01\rangle \otimes (\alpha|1\rangle + \beta|0\rangle) + \frac{1}{2}|10\rangle \otimes (\alpha|0\rangle - \beta|1\rangle) + \frac{1}{2}|11\rangle \otimes (\alpha|1\rangle - \beta|0\rangle)$$

$$\tag{5}$$

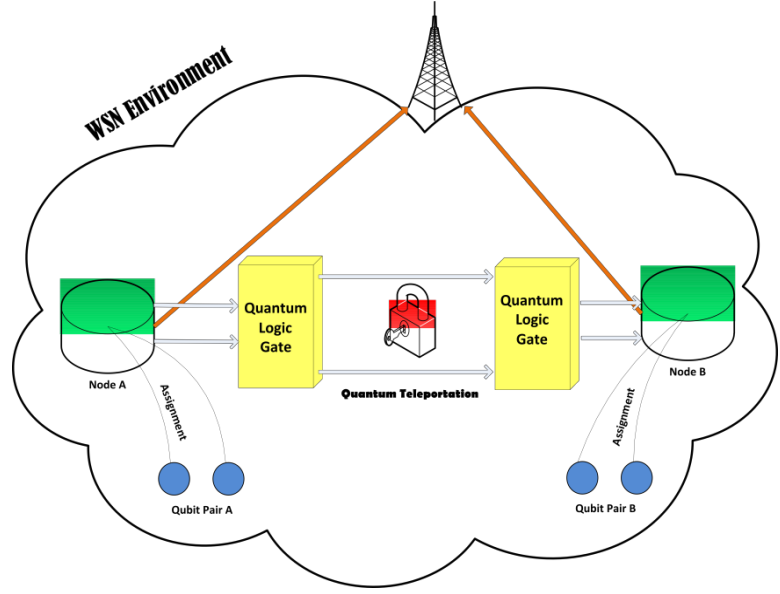

Fig. 3. Security of WSN using quantum logic gates

This equation (5) expresses the possible outcomes of measuring the first two qubits. The state of the third qubit transforms into one of the four states, $\alpha|0\rangle + \beta|1\rangle$, $\alpha|1\rangle + \beta|0\rangle$, $\alpha|0\rangle - \beta|1\rangle$, $\alpha|1\rangle - \beta|0\rangle$ corresponding to the measurements of the first two qubits namely $|100\rangle, |101\rangle, |110\rangle, |111\rangle$. Thus, the receiver can change the state of the third qubit to $|\phi\rangle$ by applying the appropriate operations, according to the measurements of the first two qubits received from the sender. Finally, the quantum state is teleported from the sender to the receiver.

Fig.3 shows a wireless sensor network environment in which two nodes A and B are communicating to a remote base-station. Each node is assigned a qubit pair. The two nodes are communicating with each other through a mechanism of quantum teleportation using quantum logic gates. The use of the quantum logic gates makes the environment a more reliable, secure and less susceptible to malicious attacks.

To make the system more secure, we can use a 3x3 Toffoli gate as well for this purpose.

Again consider a quantum state $|\phi_0\rangle$ is initially prepared as

$$|\phi_0\rangle = |\phi\rangle \otimes |0\rangle \otimes |0\rangle$$

$$|\phi_0\rangle = (\alpha|0\rangle + \beta|1\rangle) \otimes |00\rangle$$

$$|\phi_0\rangle = \alpha|000\rangle + \beta|100\rangle \quad (6)$$

Applying a Hadamard gate is applied to the second qubit to obtain quantum state $|\phi_1\rangle$ as:

$$|\phi_1\rangle = \alpha|0\rangle \otimes \frac{1}{\sqrt{2}}(|0\rangle + |1\rangle) \otimes |0\rangle + \beta|1\rangle \otimes \frac{1}{\sqrt{2}}(|0\rangle + |1\rangle) \otimes |0\rangle$$

$$|\phi_1\rangle = \frac{\alpha}{\sqrt{2}}(|000\rangle + |010\rangle) + \frac{\beta}{\sqrt{2}}(|101\rangle + |110\rangle) \quad (7)$$

Next, applying 3x3 Toffoli gate from second to the third qubit we get equation (8)

$$|\phi_2\rangle = \frac{\alpha}{\sqrt{2}}(TG|000\rangle) + TG|010\rangle) + \frac{\beta}{\sqrt{2}}(TG|101\rangle) + TG|110\rangle)$$

$$|\phi_2\rangle = \frac{\alpha}{\sqrt{2}}(|001\rangle + |010\rangle) + \frac{\beta}{\sqrt{2}}(|101\rangle + |110\rangle) \quad (8)$$

If now we apply the Toffoli gate to $|\dot\phi_2\rangle$, we generate $|\dot\phi_3\rangle$

$$|\phi_3\rangle = \frac{\alpha}{\sqrt{2}}(|000\rangle + |010\rangle) + \frac{\beta}{\sqrt{2}}(|101\rangle + |110\rangle) \quad (9)$$

Applying Hadamard gate to the first qubit it produces $|\dot\phi_4\rangle$ as

$$|\phi_4\rangle = \frac{\alpha}{\sqrt{2}} \times \frac{1}{\sqrt{2}}(|0\rangle + |1\rangle) \otimes (|00\rangle + |10\rangle) + \frac{\beta}{\sqrt{2}} \times \frac{1}{\sqrt{2}}(|0\rangle - |1\rangle) \otimes (|01\rangle + |10\rangle)$$

$$|\phi_4\rangle = \frac{1}{2}|00\rangle \otimes (\alpha|0\rangle + \beta|1\rangle) + \frac{1}{2}|01\rangle(\alpha|1\rangle + \beta|0\rangle) + \frac{1}{2}|10\rangle \otimes (\alpha|0\rangle - \beta|1\rangle) + \frac{1}{2}|11\rangle(\alpha|1\rangle - \beta|0\rangle) \quad (10)$$

which is generating the same results as of equation (5)

## IV. CONCLUSION AND FUTURE WORK

In this research, the authors have tried achieving a secure and reliable large scale wireless network requiring efficient key distribution and management mechanisms; quantum key distribution offering huge advantages over traditional methods by the use of entanglement swapping and quantum teleportation; adoption of a EPR-pair allocation scheme in terms of quantum gates like Toffoli and Feynman to overcome the susceptibility caused by malicious nodes. The mathematical results clearly indicate that whichever gate is used at intermediate stages we can extract the same information at the output as is generated at the input and intermediate communication can be made more secure with the use of different types of quantum gates. We have shown this for two different types of quantum gates, one with the 2x2 gate and the other with the 3x3 gate. This can be extended to 4x4 gates as well. Quantum cryptography is a promising domain and much contribution in various applications of this area are in progress and quite challenging.

.